\documentclass[twocolumn,showpacs,preprintnumbers,amsmath,amssymb]{revtex4}
\usepackage{graphicx}% Include figure files
\usepackage{dcolumn}% Align table columns on decimal point
\usepackage{bm}% bold math

\def\keV{\,{\rm keV}}

\begin{document}

\preprint{APS/123-QED}

\title{Searching for decaying axion-like dark matter from clusters of galaxies}
% Force line breaks with \\

\author{Signe Riemer-S\o rensen$^1$} \email{signe@dark-cosmology.dk}
\author{Konstantin Zioutas$^{2}$} \author{Steen H. Hansen$^1$}
\author{Kristian Pedersen$^1$} \author{H\aa kon Dahle$^{1,3}$}
\author{Anastasios Liolios$^4$}
 \affiliation{ $^1$Dark Cosmology Centre, Niels Bohr Institute, University of Copenhagen, Juliane Maries Vej 30, DK-2100 Copenhagen, Denmark\\
$^2$Department of Physics, University of Patras, Patras, Greece and CERN, Geneva, Switzerland\\
%$^3$CERN, Geneva, Switzerland\\ 
$^3$Institute of Theoretical Astrophysics, University of Oslo, P.O. Box 1029, Blindern, N-0315 Oslo, Norway \\ 
$^4$Physics Department, Aristotle University of Thessaloniki, GR-54124 Thessaloniki, Greece}

\date{\today}% It is always \today, today,

\begin{abstract}
We consider the possibility of constraining the lifetime of radiatively decaying dark matter particles in clusters of galaxies inspired from generic axions of the Kaluza-Klein type. Such axions have been invoked as a possible explanation for the
coronal X-ray emission from the Sun. These axions, or similar particles, can be produced inside stars and some of them remain confined by the deep gravitational potential of clusters of galaxies. Specifically, we consider regions within merging galaxy clusters (Abell~520 and the ``Bullet Cluster''), where gravitational lensing observations have identified massive, but baryon poor, 
structures. From an analysis of X-ray observations of these mass concentrations, and the expected photon spectrum of decaying solar KK-axions, we derive lower limits to the lifetime of such axions of $\tau \approx 10^{23}$~sec. 
However, if KK-axions constitute less than a few percent of the dark matter mass, this lifetime constraint is similar to that derived from solar KK-axions.
\end{abstract}

%Use showkeys class option if keyword
\maketitle
\section{Introduction}
The particle nature of the dark matter remains a mystery. Whereas cosmological observations have determined that there is approximately six times as much dark matter as there is baryonic matter, we still have not identified which particle or particles constitute the dark matter.

An apparently unrelated issue is "the solar corona problem": In order to explain the quiet solar X-ray spectrum, one may need to invoke new physics, such as introducing massive axions of the Kaluza-Klein (KK) type \cite{DiLella-Zioutas,2006:Asztalos}. The ÒstandardÓ axions with the remaining rest mass window in the sub-eV range \cite{Hannestad:2005} live far too long to solve this problem. KK-axions appear in the framework of large extra dimensions \cite{ADD,HKL} but they are not considered necessarily to be dark matter candidates. In this work a generic example of relatively massive radiatively decaying axions has been used. In the large extra dimensions scenario, it is assumed that only gravity propagates in the higher-dimensional space, while the standard-model fields are confined to our (3+1)-dimensional subspace. Since axions are singlets under the standard-model gauge group, they could also propagate in the higher-dimensional space. As a result of compactification, the higher-dimensional axion field is decomposed into a KK tower of states with the mass spacing of order $1/R$, where R is the compactification radius of the extra dimensions \cite{HKL}. All KK excitations have the same coupling strength to matter. A source of axions will emit all KK states up to the kinematic limit. 
Another interesting feature of the higher-dimensional theory is that the axion mass $m_a$, corresponding to the lowest KK state, may decouple from the PQ scale and may be determined by the compactification radius, $m_a = 1/(2R)$.
% Could be removed when shortening!

We will here address the possibility that part of the dark matter in galaxy clusters are like KK-axions which are produced in normal stars and accumulated near their birth place over the lifetimes of these stars \cite{DiLella-Zioutas,HR2002,HR2003}. Part of the so produced axions are confined by the gravitational potential of the host galaxy clusters, where they stay until they eventually decay spontaneously. These KK-axions are predicted to decay into two photons, producing a broad characteristic spectrum which happens to peak just where current astrophysical X-ray instruments are most sensitive, i.e. in the few keV range.

We will consider X-ray observations of merging clusters, from which weak gravitational lensing observations have provided strong evidence for a  matter component which is non-baryonic. Assuming that parts of this dark matter is of the KK-axion type, we can derive the first cosmological constraints on the lifetime of this kind of dark matter particles.

\section{Radiatively decaying particles}\label{sec:axion_spectrum}
\begin{figure}
	\includegraphics[angle=90,width=8.6cm]{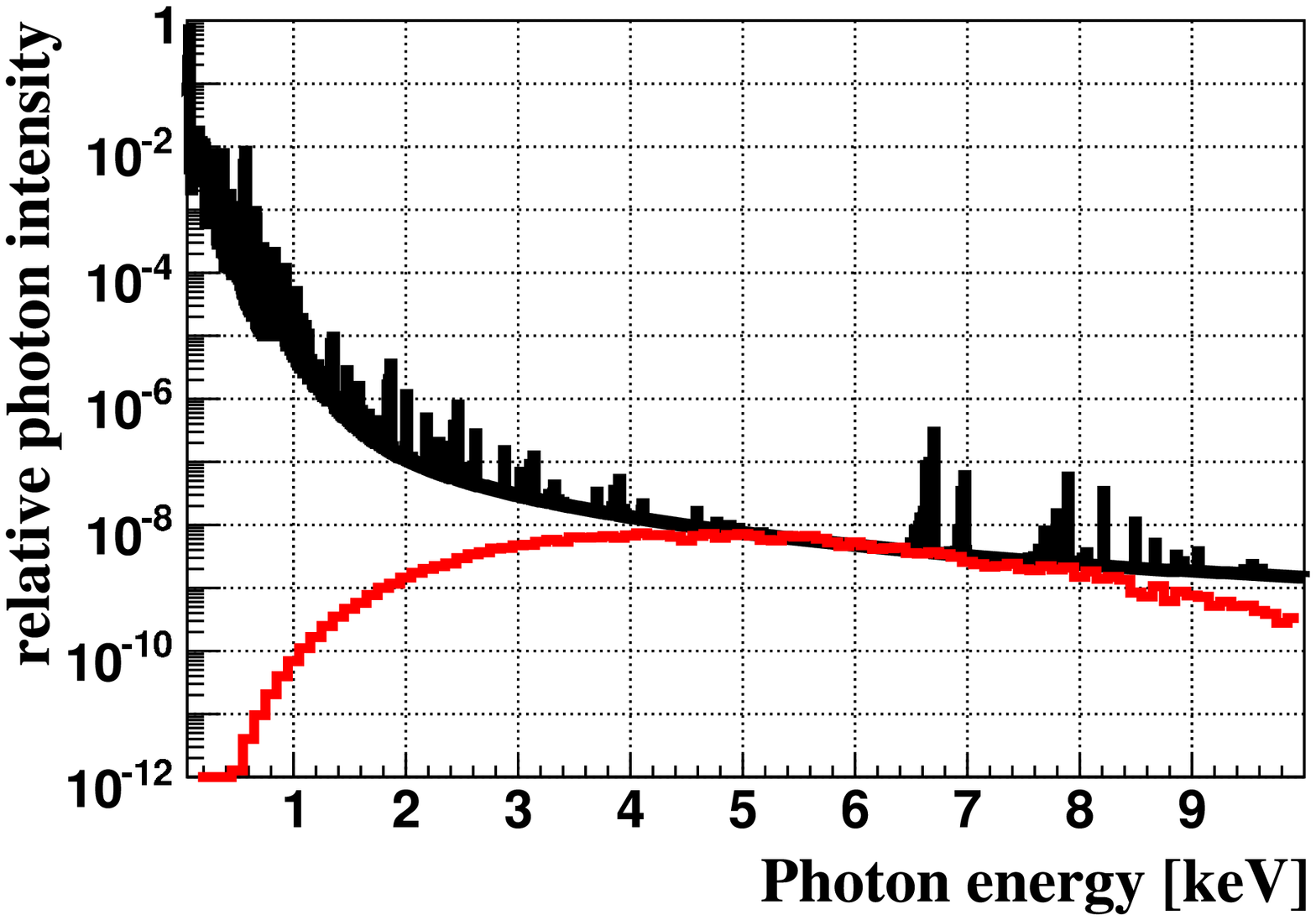}
	\caption{The reconstructed quiet Sun X-ray spectrum (black, upper curve) \cite{Reale:2001}
	and the expected flux from decaying massive axions of the Kaluza-Klein type (red, lower curve) \cite{DiLella-Zioutas}.}
	\label{fig:expected_flux}
\end{figure}

All of the massive KK-axions produced in the Sun as a possible explanation for the quiet solar X-ray spectrum \cite{DiLella-Zioutas}, can decay radiatively, and from the assumed production mechanism, we can calculate the resulting photon spectrum. For solar KK-axions \cite{DiLella-Zioutas} the peak energy lies at approximately $4.5$~keV (red line in Figure \ref{fig:expected_flux}), and for $E_\gamma \geq 2.0$~keV the derived spectrum is well represented by the following expression:
\begin{eqnarray} \label{eqn:exp_flux}
%F_{der} (E_\gamma < 2.0 \keV) &=& 0.096 E_\gamma^{-2.7} e^{-(2.1/E_\gamma)}\\
F_{der} (E_\gamma>2.0\keV) &=& 2.2\times10^{8} E_\gamma^{-8.1} e^{-(32.0/E_\gamma)}  %\, .
\end{eqnarray}

The shape and the energy of the maximum of the spectrum are tightly bound by solar physics processes. The normalization of the spectrum depends on the coupling of the KK-axions and can essentially be treated as a free parameter. Assuming the Sun to be a ``typical'' star in cosmos, which is a reasonable assumption, Equation \ref{eqn:exp_flux} can be applied to KK-axions produced in stars and confined by the gravitational dark matter well of clusters of galaxies.
Solar X-ray observations made with the YOHKOH mission fit the general feature of the radial distribution of the massive solar axion model, except at low photon energies (below ~4 keV), where the bulk of the quiet and not-quiet solar X-ray spectrum is emitted (see black line of Figure 1, and discussion in ref. [10]). 

\section{Observed clusters of galaxies}
Observations with the {\it Chandra X-ray Observatory} of the two clusters of galaxies Abell~520 and the ``Bullet Cluster'' (also known as 1E0657-558) were analyzed. Basic data on the clusters and the observations are specified in Table 1 for a cosmology with $H_0=0.71$~km/s/Mpc, $\Omega_M=0.3$, $\Omega_\Lambda=0.7$ \cite{Spergel:2006}. The data were retrieved from
the public Chandra archive and processed using standard data reduction methods with CIAO version 3.3 \cite{ciao}.

\begin{figure}	
	\includegraphics[width=8.6cm, height=8.6cm]{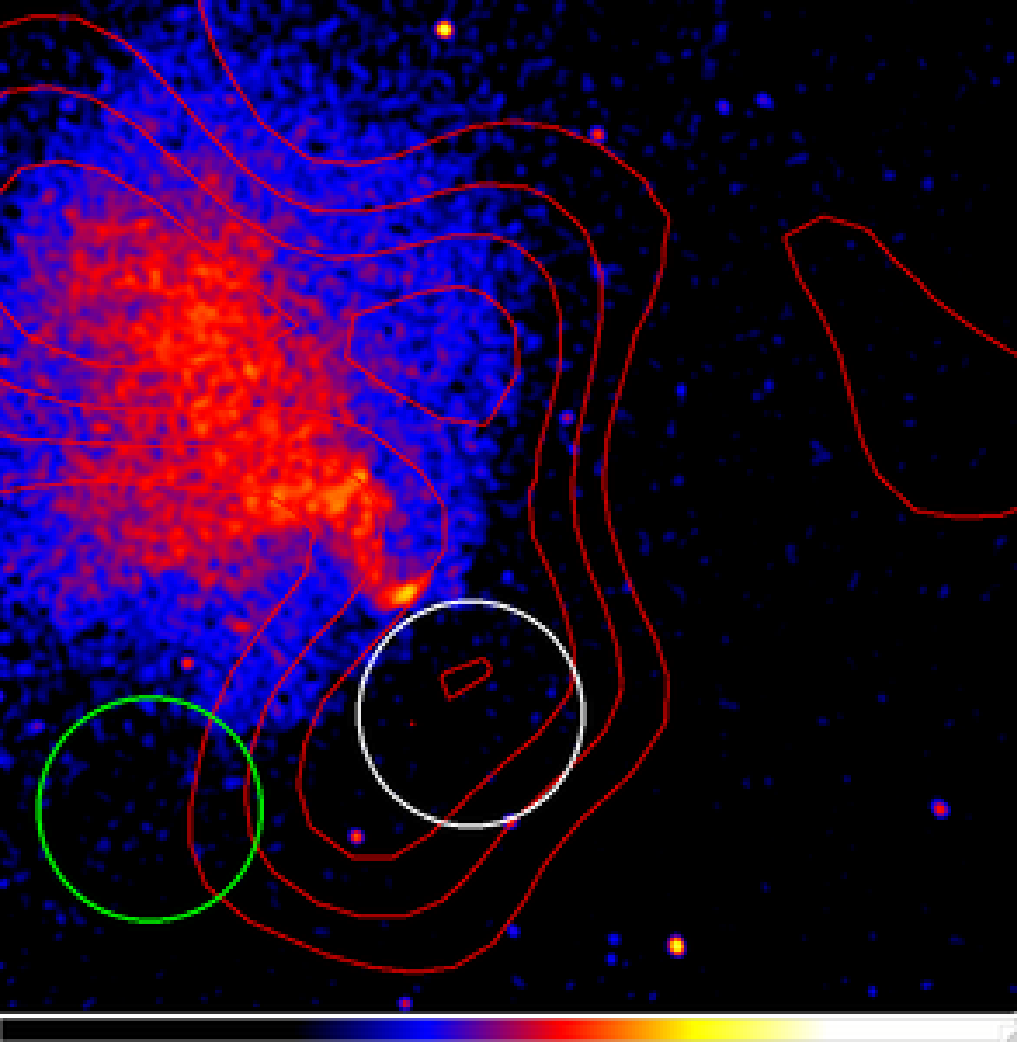}
         \includegraphics[width=8.6cm, height=8.6cm]{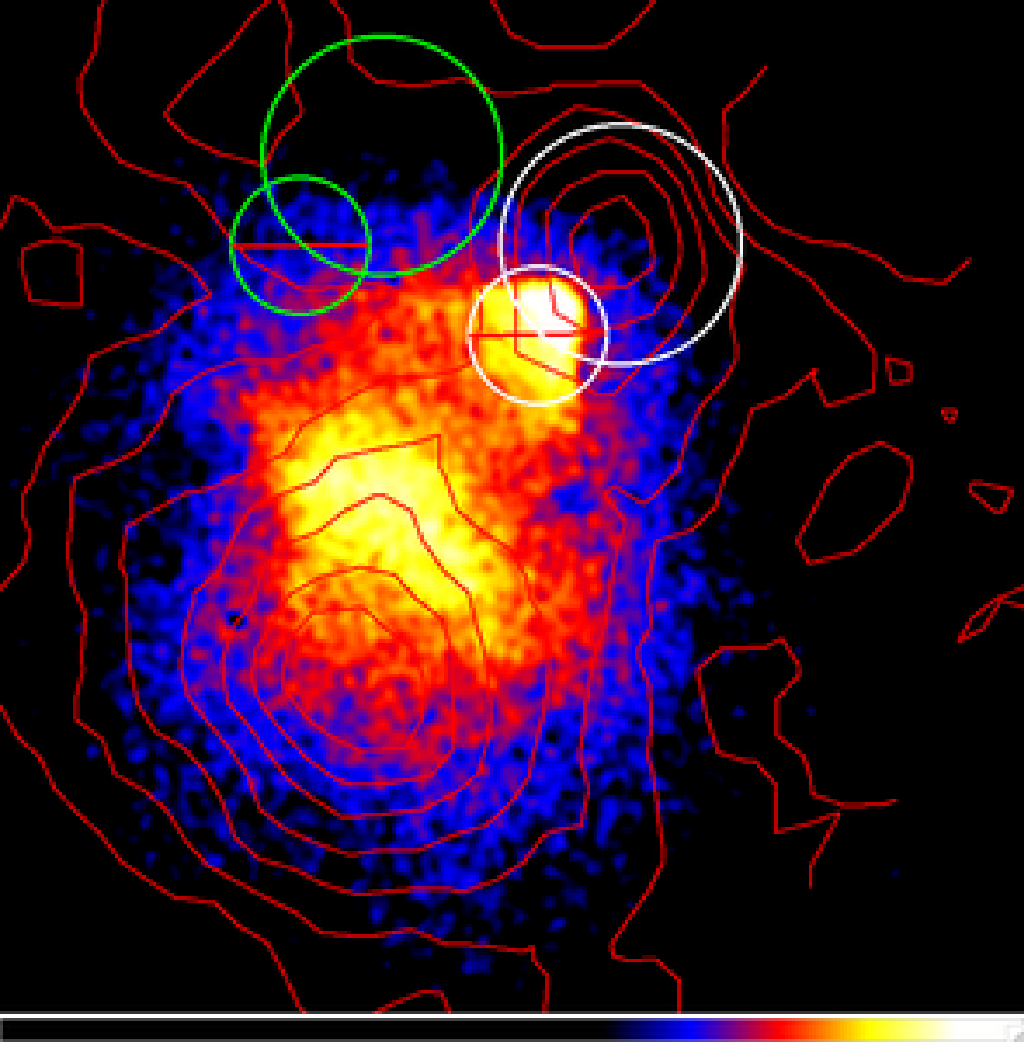}
	\caption{The X-ray image of Abell~520 (above) and the Bullet Cluster (below) overlaid the
	gravitational potential from weak gravitational lensing (red, Bullet Cluster contours from \cite{Clowe:2006}) and the dark matter blob region (white), and the reference region (green).}
	\label{fig:regions}
\end{figure}

As seen from the X-ray images in Figure \ref{fig:regions} both clusters are merging systems and the Bullet Cluster is showing a prominent bow shock feature \cite{Markevitch:2004,Clowe:2006}. In such disturbed systems the hot X-ray emitting intracluster plasma (which is the dominating baryonic component in clusters) is displaced from the stars and the dark matter, leaving mass concentrations practically devoid of baryons. These so-called ``dark matter blobs'' are ideal environments for dark matter studies \cite{Clowe:2006,Boyarsky-bullet,A520} since they have a high dark matter density and a low contamination of X-ray ``noise'' from baryonic matter.
%\begin{figure*}	
%	\caption{The X-ray image of Abell~520 (left) and the Bullet Cluster (right) overlaid the gravitational potential from weak lensing (red, Bullet Cluster contours from \cite{Clowe:2006}) and the blob region (white), and the reference region (green).}

%\end{figure*}

\section{Data analysis} \label{sec:data_analysis}
For both clusters, two regions are selected for the analysis. One region is covering a dark matter blob with high mass but low X-ray emission (white in Figure \ref{fig:regions}). The other region is a reference region with the same size and form and of similar X-ray flux, but much smaller mass (green in Figure \ref{fig:regions}). For comparison to the results of \cite{Boyarsky-bullet} the region covering the dark matter blob of the Bullet Cluster is chosen identical to their SUB region i.e. as the large circle centered on the mass peak with the smaller circle centered on the shock front subtractred. For both clusters, there is a mass contrast of an order of magnitude between the dark matter blob region and the reference region in which the baryons are the dominating source of X-ray emission.

The dark matter in the Milky Way halo also contributes to the signal \cite{Riemer-Sorensen:2006fh}, and we must first remove such a contamination. Hence, before extracting the spectra, a background region with small mass and no significant X-ray emission, is chosen from the same observation and subtracted. In this way any possible signal from the galactic halo is subtracted, simplifying the expected signal from the decaying dark matter as only one source redshift has to be considered. 
However, the X-ray flux from the Milky Way (which is proportional to the ratio of the Milky Way halo mass within the considered fields of view divided by the mean halo distance squared) is negligible compared to the cluster X-ray flux, i.e. $M_{halo}/D_{halo}^2 << M_{blob}/D_{blob}^2$. 
% Can be removed for shortening

\begin{figure}
	\includegraphics[angle=270,width=8.6cm]{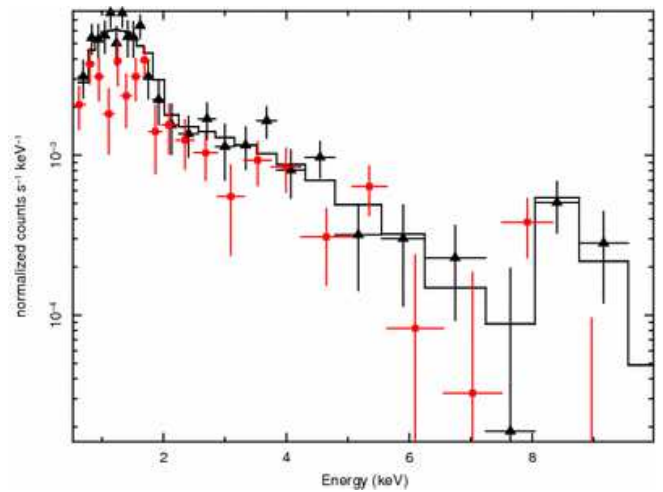}
	\caption{The observed spectrum of the blob region (red, squares) of Abell~520 and the spectrum of the reference region (black, triangles) with the fitted ``basis model'' (black, solid, reduced $\chi^2 \approx 1$).}
	\label{fig:spectrum_ref}
\end{figure}

A convenient ``basis model'' consisting of a thermal plasma model and several gaussians is fitted to the baryon dominated non-blob reference region as shown in Figure \ref{fig:spectrum_ref} for Abell~520. No physical quantities are derived from this model, it is just designed to fit the data with a reduced $\chi^2\approx1$ for the numerical analysis of the spectrum (however, the resulting physical parameters are typical for clusters). All these baryonic model parameters, except the normalization, are frozen and the model describing the axion flux is added (the red line in figure \ref{fig:expected_flux}). The continuous spectrum of decaying axions is represented by redshifting the analytic expression in Equation \ref{eqn:exp_flux} according to the distance of the cluster of galaxies.
%a redshifted powerlaw with an exponential absorbtion as explained in Section \ref{sec:axion_spectrum}. 
The composite model is fitted to the $2.0-9.5$~keV spectrum of the dark matter blob region with the normalization of the two components as the only free parameters. There are no outstanding differences between the spectrum from the dark matter blob region and the baryon dominated reference region.

An upper limit on the X-ray flux from the dark matter blob region is determined using two different methods. In the first method, all of the flux received from the region is taken as a conservative upper limit on the flux originating from decaying dark matter. This method is independent of the assumed form of the decaying dark matter spectrum. The $3\sigma$ upper limit (with respect to the fitting residuals) on the flux is determined from the fit with the normalization given by the fitted value plus $3\sigma$. The obtained upper limits on the dark matter related X-ray luminosities are given in Table 1. In the second method, the decaying dark matter spectrum is assumed to have the form shown in Figure \ref{fig:expected_flux} (red line). An upper limit on the X-ray flux from a spectrum with this form which can be added to the ``basis model'', is determined by increasing the derived model normalization systematically in small steps while leaving the normalization of the ``basis model'' as a free parameter. When the reduced $\chi^2$ reaches a value of $10$ ($\Delta\chi^2=9$), the $3\sigma$ upper bound on the luminosity of the axion spectrum alone is determined.

\section{Mass determinations} \label{sec:mass}
For Abell~520 the masses of the regions were determined from weak gravitational lensing. The values are based on measuring the overdensity in the regions with respect to the mean density in a surrounding annulus with inner and outer radius of 0.85 arcmin and 4 arcmin, respectively. Hence, the mass value can be regarded as a conservative lower limit on the mass contained within the region. A detailed description of the data and methodology of the weak lensing analysis is given elsewhere \cite{Dahle:2002}. The mass of the Bullet Cluster blob region (SUB) is taken from \cite{Boyarsky-bullet}.

The gas masses of the regions can be estimated assuming a geometry for the gas distribution, e.g. that the gas is spherically distributed and with an average density derived from the thermal plasma model (MEKAL \cite{mekal}) fitted to the spectra. As seen in Table 1, the gas masses of the blobs are much lower than their total masses (in agreement with the generally observed gas mass fraction of $f_{gas}\approx0.11$ \cite{Allen:2002}). Hence, the gas masses can be neglected and the mass of the dark matter is taken to be the total mass of the blobs determined from gravitational lensing.

\begin{table*}
\begin{ruledtabular}
\begin{tabular}{lcc|cc}
								& \multicolumn{2}{c|}{{\bf Abell~520}} & \multicolumn{2}{c}{{\bf Bullet Cluster}}\\
								& Value & Ref. & Value & Ref\\ \hline
Redshift                      					&  0.2 & \cite{Ebeling:1998} & 0.29 & \cite{Boyarsky-bullet} \\
$D_A$, [Mpc]         					&  662 & & 872 	& \\
Region radius, $\delta \theta$ [arcmin] 	&  0.85 & & 0.66 & \\   
{\it Chandra} observation id						& 4215 & \cite{heasarc} & 5356 and 5361 & \cite{heasarc} \\
Exposure time [ksec]					& 67	& & 179 & \\
\hline
Blob region values                   			& & & & \\ \hline 
$M_{total}$ [$10^{13}M_\odot$]  		& $6.7\pm2.1$ & & $5.8$ & \cite{Boyarsky-bullet}  \\
$M_{gas}$ [$10^{13}M_\odot$]    		& $0.4$ & & $0.7$ & \cite{Boyarsky-bullet}  \\
Basis model, reduced $\chi^2$ reference region & $0.75$ & & $1.05$ &  \\
Basis model, reduced $\chi^2$ blob region & $0.78$ & & $1.04$ & \\
Basis + expected, reduced $\chi^2$ 	& $0.77$ & & $1.05$ &\\
$3\sigma$ total luminosity upper limit [$10^{44}$erg/sec] & $0.2$ & & $0.9$ &\\
$3\sigma$ form dependent luminosity upper limit [$10^{44}$erg/sec] & $0.2$ & & $1.4$ & \\
\end{tabular}
\end{ruledtabular}
\caption{The obtained values as described in Section \ref{sec:data_analysis} and \ref{sec:mass} for a cosmology with $h=0.71$, $\Omega_M=0.3$, $\Omega_\Lambda=0.7$ \cite{Spergel:2006}.}
\end{table*}

\section{Lifetimes}
A simple estimate of the lifetime of the dark matter particles can be derived from the observed luminosities: $L=\Gamma E_\gamma N$, where $\Gamma$ is the decay rate, $E_\gamma$ is the photon energy, and $N$ is the total number of particles. In the general case of non-relativistic radiative dark matter two-body decays, $E_\gamma = m/2$ for a dark matter candidate of mass $m$, $N=X M_{DM}/m$, where $X$ is the mass fraction of the dark matter made up of the considered candidate. 
It is worth mentioning that we are considering the case where the many individual decay modes
produce a wide KK-axion bump, which has the shape of the red (lower) curve in figure
\ref{fig:expected_flux}. 
The lifetime then becomes:
\begin{equation}
\tau=\frac{1}{\Gamma}=\frac{2XM_{DM}}{L}
\end{equation}
For the two clusters the observational $3\sigma$ upper limits on the luminosities lead to a lower limit on the mean lifetime of the order of $\tau \gtrsim 10^{23}$~sec. The strongest constraint comes from Abell~520 and gives an upper limit of $\tau \gtrsim 6\times10^{24}$~sec assuming all of the dark matter to be made up of one single candidate with a radiative two-body decay. Of course the assumed axions produced in the stars (as predicted for the Sun), do not necessarily have to constitute all of the dark matter in the clusters. Still they would be confined by the gravitational potential of the cluster dark matter and an enhanced signal from decaying massive axions or the like would be expected from dark matter dense regions (such as the blobs). If this type of axions for instance contribute to $2\%$ of the total dark matter mass, the lifetime constraint relaxes to $\tau\gtrsim 10^{23}$~sec (which corresponds to $g_{a\gamma\gamma} \lesssim 3\times 10^{-15}$~GeV$^{-1}$ for a mean axion rest mass of 5~keV). This first rough estimate of the lifetime is actually not significantly different from that derived for solar KK-axions of $\tau \approx 1.25\times 10^{20}$~sec for a mean axion rest mass of 5~keV \cite{DiLella-Zioutas}.

Moreover, we compare the cosmic densities of energy associated with dark matter and electromagnetic radiation \cite{Fukugita:2004} since it will give a crude estimate of the radiative lifetime of dark matter particles on cosmic scale. We assume that the decay of a single kind of dark matter particles is responsible for the main part of the present-day electromagnetic radiation of the Universe. The total fraction of post-stellar radiation densities from resolved sources in radio-microwave, far infrared, optical, and X-$\gamma$-rays is $d_{tot} \approx 2.4 \times 10^{-6}$ \cite{Fukugita:2004}. With the age of the Universe taken to be $\tau_{Universe} = 4.5 \times 10^{17}$~sec \cite{Spergel:2006}, this gives an order of magnitude estimate of the lifetime, independent of particle type, of:
\begin{equation}
\tau \approx \frac{\Omega_{DM}\tau_{Universe}}{d_{tot}}\approx4\times10^{22} \textrm{sec} \,,
\end{equation}
which is similar to the life times derived above. It is encouraging that studying KK-axion decay on scales from the Sun, to clusters of galaxies, to the size of the observable Universe results in life time constraints of similar order-of-magnitude.

\section{Summary}
Dark matter blobs in merging clusters of galaxies are excellent laboratories for constraining fundamental properties of dark matter candidates, like their lifetime. One promising dark matter constituent might have properties similar to the generic KK-axion, presumably produced inside stars and proposed to explain the origin of the quiet solar X-ray spectrum which has remained elusive for several decades. Trapped by the deep gravitational potential of clusters of galaxies some of the KK-axions, or the like, eventually decay isotropically into two X-ray photons and thereby contribute to the diffuse intracluster X-ray emission. Hence, the X-ray emission from dark matter blobs can be taken as a conservative upper limit on the assumed total KK-axion luminosity, leading to a lower limit on the KK-axion lifetime of  $\tau \gtrsim 10^{23}$~sec. This is several orders of magnitude larger than the lifetime derived for solar KK-axions ($10^{20}$~sec). However, if KK-axions constitute less than a few percent of the dark matter mass, the two life time constraints are similar.

\begin{acknowledgments}
The authors would like to thank S.~Hannestad and B.~Lakic for useful comments. KZ thanks the Dark Cosmology Centre for the support and the warm hospitality, where the first stage of this work was performed. The Dark Cosmology Centre is funded by the Danish National Research Foundation. AL and KZ acknowledge support by the EU-network on direct dark matter detection of the ILIAS integrating activity (contract number: RII3-CT-2003-506222).
\end{acknowledgments}

\end{document}